\documentclass[10pt]{iopart}
\usepackage{amsfonts}

\def\bm#1{\mbox{\boldmath{$#1$}}}

\def\P{{\cal P}}

\def\p{\partial}
\def\a{\alpha}

\def\d{\delta}

\def\f{\varphi}

\def\r{\rho}
\def\g{\gamma}
\def\G{\Gamma}

\def\ra{\rightarrow}

\newcommand{\be}{\begin{equation}}
\newcommand{\ee}{\end{equation}}
\newcommand{\bea}{\begin{eqnarray}}
\newcommand{\eea}{\end{eqnarray}}

\begin{document}

\title{Scaling Laws in the Cosmic Structure and Renormalization Group}

\author{Jos\'e Gaite\dag\ and Alvaro Dom{\'\i}nguez\ddag}

\address{\dag\ IMAFF,
CSIC, Serrano 113 bis, E-28006 Madrid, Spain
}

\address{\ddag\ F{\'\i}sica Te\'orica, Univ.\ Sevilla, Apdo.\ 1065,
  E-41080, Sevilla, Spain}

\begin{abstract}
  There is evidence of a scale-invariant matter distribution up to
  scales over 10 Megaparsecs.  We review scaling (fractal or
  multifractal) models of large scale structure and their
  observational evidence.  We conclude that the dynamics of
  cosmological structure formation seems to be driven to a
  multifractal attractor. This supports previous studies, which we
  review, of structure formation by means of the renormalization group
  within a hydrodynamic formulation.
\end{abstract}

\section{Introduction}
Both the old idea of hierarchical clustering in the Universe and the
statistical analysis of the distribution of galaxies have led to
scaling laws in the cosmic structure and fractal models of it
\cite{Piet}.  Scaling is usually demonstrated by the appearance of
power laws in the correlation functions.  There are other scaling laws
in the cosmic structure, referring to other features.  For example,
the counterpart of galaxy clusters are galaxy {\em voids}, namely,
large empty regions in the galaxy distribution.  Fractal voids have
scaling properties in the rank-ordering of voids \cite{myvoids},
verified in galaxy surveys \cite{Rus}.

Nevertheless, a consensus on the range of application of scaling laws in the
description of the cosmic structure has not been reached (a recent discussion
is in \cite{Jones-RMP}).  The most general scaling model is the multifractal
model, introduced in cosmology to describe ``non-uniform'' fractal clustering
\cite{MF}. However, the study of cosmological $N$-body simulations has led to
halo models of large scale structure \cite{CoorShet} which do not assume any
scaling laws. Fortunately, multifractal models can be formulated in a way that
closely resembles usual halo models, allowing us to unveil scaling laws in
$N$-body simulations \cite{EPL}.

Further to the description of cosmic structure, the problem of
structure formation has given rise to scaling laws with dynamical
content. For example, a popular model of structure formation, the
adhesion model \cite{adhesion,ShZe89}, displays dynamical scaling and,
in addition, leads to multifractal structure \cite{V-Frisch}.
Structure formation is a nonlinear process, like other {\em
non-equilibrium} processes in statistical mechanics that have scale
invariance. Scaling is crucial in critical phenomena, which are
phenomena that take place in equilibrium statistical mechanics for
definite values of the parameters.  The renormalization group (RG),
which arose in quantum field theory, was soon applied to critical
phenomena \cite{WilKo} and, more recently, it has also been applied to
non-equilibrium processes (``dynamical RG'').  The latter application
can be extended to the process of structure formation in
cosmology. However, the dynamical RG approach is perturbative and
encounters some problems in this regard.

A basic feature of modern formulations of the RG is that they deal
with many-body systems by progressively removing irrelevant degrees of
freedom until the ``renormalized system'' becomes manageable.  The
coarse-graining procedure, widely used in statistical mechanics and
hydrodynamics, is inspired in the same idea.  It is intrinsically
non-perturbative and can be applied to structure formation.

We begin by reviewing evidence of scaling in the large scale structure
of matter, arising from observational data or cosmological $N$-body
simulations.  Our analysis is consistent with the hypothesis that the
dynamics of structure formation is driven to a {\em multifractal
attractor}, like other non-equilibrium processes in statistical
mechanics. This result justifies the hypothesis of {\em dynamical}
scaling and, in particular, the dynamical RG approach.  We connect
with work along this line within the adhesion model, which is quite
succesful in one dimension, but seems to require additional
ingredients in three dimensions.  Therefore, we turn to approaches
based on the coarse-graining procedure and in particular, we formulate
the coarse-graining ``exact'' RG group for structure formation.

Since we assume the existence of a multifractal attractor, we will not
review approaches based on perturbation theory about the linear
regime.  Here we just mention the interesting recent application of
the RG by Crocce \& Scoccimarro \cite{Scoc}.

\section{Scaling laws in the large scale structure}

\subsection{Scaling of galaxy clustering}
\label{sec:galaxies}

Hierarchical clustering consists of a hierarchy of clusters of clusters.
In general, fractal geometry studies sets (or functions) that are irregular
(non-smooth) and have {\em fine structure}, namely, detail at all scales.
Usually, the fine structure of a fractal is due to its {\em self-similarity},
that is, to the set being similar to parts of itself, in a strict or
approximate sense. Random fractals only have {\em statistical}
self-similarity, which implies that the correlation functions are power laws.

A useful description of random fractals is through the {\em
number-radius} relation, which expresses the number of points in a
ball of radius $r$ centered on one point and {\em averaged} over every
point: it has to be the power law $N(r) = B\,r^D$, where $D$ is the
fractal dimension and $B$ a constant.  $N(r)$ is the cumulative
conditional probability, that is, the integral of the conditional
probability $\G(r)$, which measures the average probability of finding
another point at distance $r$ from one given point.  In turn, $\G(r)$
is directly related to the reduced two-point correlation function
$\xi(r)$, namely, $\G(r) = \bar{\r}\,\left(1+\xi(r)\right)$, where
$\bar{\r}$ is the average density.  Both $N(r)$ and $\G(r)$ are used
to test scaling: their log-log plots must be linear, with slopes, $D$
and $D-3$, respectively.

The distribution must have a transition to homogeneity on very large scales,
where $D \ra 3$. The corresponding scale can be defined in terms of the
non-dimensional correlation $\xi(r)$, by writing it as $\xi(r) =
(r_0/r)^\g$, where $\g = 3-D$ and $r_0$ is the scale of transition to
homogeneity. We see that the strongly non-linear regime, $\xi \gg 1$, is the
fractal regime, where $\xi$, $\G$ {\em and} $N$ are all power laws.  In the
homogeneous regime, for $r \gg r_0$, $\xi \ra 0$ and $\G \ra \bar{\r}$, while
$N \propto r^3$.

Typical values of the fractal dimension and the scale of transition to
homogeneity are about $D \simeq 2$ \cite{Piet,Tikho} and $r_0 \simeq 15$ Mpc
$h^{-1}$ \cite{Tikho}, respectively.

\subsubsection{Scaling of voids}

Voids in the galaxy distribution scale if the number of voids with a
given size is a power law of the size. It is again convenient to
employ the cumulative count, namely, the number of voids $N(L > \ell)$
with linear size larger than a reference $\ell$, which fulfills $N(L >
\ell) \propto \ell^{-D}$, where $D$ is the fractal dimension
\cite{myvoids}. The cumulative count is the {\em rank}, so the
preceding law can also be expressed as a power-law dependence of size
with rank (if size refers to volume, the exponent is $3/D$).  Such
type of dependence is an instance of {\em Zipf's law}.

The scaling of voids in galaxy surveys is still uncertain. Recent
analyses find sets of convex-like voids that satisfy Zipf's law
\cite{Rus}.  The fractal dimension deduced from them, $D \simeq 2$,
coincides with the previously cited dimension deduced from clustering.
However, $D = 2$ is the (box-counting) dimension of the boundary of
voids and, therefore, the minimal value of $D$ in the Zipf law for
voids \cite{cut-out}. The actual fractal dimensions of the samples may
be smaller \cite{Rus}.

\subsubsection{Luminosity segregation}

Different galaxy populations may have different statistical
properties. If these populations are fractal, they may have different
dimensions.  In fact, although $D \simeq 2$ is typical, other analyses
yield smaller values, and one can change $D$ somewhat by selecting
different galaxy populations: a systematic analysis of galaxy
populations in the Sloan Digital Sky Survey (SDSS) selected by
luminosity, made by M.\ Montuori \cite{Marco}, shows a decrease of
fractal dimension with luminosity.  Thus, the galaxy distribution may
be, rather than a simple fractal, a {\em multifractal}, in which
various dimensions appear naturally.

\subsection{Multifractal model}

Multifractals are the {\em most general} scaling distributions.
They appear frequently as attractors of nonlinear dynamical systems.
Multifractal measures represent highly irregular mass distributions, that is,
with mass concentrations of very different magnitude. This magnitude is
defined by the {\em local} dimension $\a(x)$:
\begin{equation}
m[B({\bm x},r)] \sim r^{\a(}\mbox{\boldmath{$^x$}}^{)},
\label{a}
\end{equation}
where $m[B({\bm x},r)]$ is the mass in the ball of radius $r$ centered on
$x$.  In a regular mass distribution, $\a = 3$ (constant), so mass
concentrations $\a({\bm x}) < 3$ are singularities.  On the other hand, an
ordinary fractal can be considered endowed with a {\em uniform} mass
distribution over it, such that $\a < 3$ is the {\em constant} fractal
dimension. Thus, in the context of multifractals, ordinary fractals
are called {\em monofractals} (or {\em unifractals}). Full-fledged
multifractals possess a range of dimensions $\a$, namely, $0 < \a_{\rm
min} \leq \a \leq \a_{\rm max}$. Every set of points in which $\a$
takes a definite value is a fractal set.  Therefore, a multifractal
can be considered as a set of interwoven fractals with running $\a$.
The {\em multifractal spectrum} $f(\a)$ is the function that gives the
dimension of the fractal with exponent $\a$.

Statistical moments are defined by
\begin{eqnarray}
M_q(r) &=& \int dm({\bm x})\; m[B({\bm x},r)]^{q-1}.
\label{Mq}
\end{eqnarray}
$M_1$ is the total mass (normalized to one).  The two-point
correlation integral $M_2(r)$ is the continuous version of the number
function $N(r)$.  Multifractals are {\em singular} non-uniform
distributions, so moments with integer $q$ are not sufficient; one has
to consider the full set of moments $M_q(r)$ for $-\infty < q <
\infty$.  We can then define the function $\tau(q)$ that gives the
scaling behaviour of this full set of moments,
\begin{equation}
M_q(r) \sim r^{\tau(q)}.
\label{tauq}
\end{equation}
$\tau(q)$ determines the multifractal spectrum through a Legendre
transform \cite{MF}: assuming $\a(q)= \tau'(q)$ to be monotone, 
$f(\a) = q(\a)\,\a - \tau[q(\a)].$

\subsubsection{The fractal distribution of halos}

We associate halos with {\em singular} mass concentrations, namely,
points with $\a({x}) < 3$, such that the density given by Eq.\
(\ref{a}) diverges as $r \ra 0$. Note that scale invariance prevents
us from assigning these singularities definite sizes or
masses. Therefore, to properly define halos we must use some small
coarse-graining scale $L$.  In $N$-body simulations, the natural
coarse-graining scale is the linear size of the volume per
particle. Initially and during the linear evolution, there is one
particle per volume element.  So halos only arise in the nonlinear
stage, as some volume elements concentrate particles from other
regions that become {\em voids}.

Therefore, we identify halos with mass concentrations of size $L$ in a
multifractal.  Since $\a \sim \log m/\log L$, every population formed
by {\em equal-mass} halos is a monofractal, although different
populations have different dimensions. We can describe this difference
between populations as a kind of {\em bias}, albeit of {\em
non-linear} type.  In a multifractal analysis of $N$-body simulations
\cite{EPL}, we have found that populations of halos of given mass are
fractals, with mass-dependent dimension.  Their common scale of
transition to homogeneity is $r_0 \simeq 14\, h^{-1}$ Mpc.

An interesting quantity is the {\em mass function} of halos, namely,
the number of halos with a given number of particles.  The
Press-Schechter formalism predicts a power law (exponentially cut off
at large mass) with exponent connected with the initial power spectrum
\cite{CoorShet}. This form is observed in our analysis, {\em but} the
power law has a fixed exponent, namely, $N(m) \sim m^{-2}$, which
corresponds to the spectral index $n=-3$ of the initial power
spectrum, just beyond the allowed range. Moreover, the initial power
spectra of the simulations we have analysed are not power laws, but
the mass funtion is always the same, {\em independently of the initial
conditions}.

\section{Dynamical scaling}

In the strongly nonlinear regime, when the initial condition is
forgotten, dynamical scaling implies that a dynamical field $\f({\bm
x},t)$ satisfies a scaling relation $\f_L - \langle \f \rangle \sim
L^{\chi} f(t/L^z)$, where $\f_L$ is the field coarse-grained over a
length $L$ (see Eq.~(\ref{window})), $\chi,\,z$ are critical
exponents, $\lim_{u \ra \infty} f(u) =1$ and $\lim_{u \ra 0} f(u) \sim
u^{\chi/z}$. In words, the fluctuations grow with time as one power
law and they eventually reach saturation, in which state they depend
on $L$ as another power law.  Dynamical scaling is customary in the
physics of surface growth and other non-equilibrium processes
\cite{Bar-Stan}.  Similarly to the situation in static critical
phenomena, the possible types of critical dynamics correspond to
(attractive) fixed points of the {\em dynamical} RG. This tool allows
one to compute the exponents $\chi$ and $z$ exactly or approximately.

\subsection{The adhesion model}

In the cosmological context, the linear regime is identified with
small departures from the homogeneus Hubble expansion. The linear
dynamics is described by a set of linearized equations for density and
velocity.  The phenomenological {\em adhesion model} is the simplest
dynamics describing nonlinear structure formation in cosmology.  It
relies on the {\em Zeldovich approximation} \cite{ShZe89}, which
consists in extrapolating into the nonlinear regime the condition of
{\em parallelism} between velocity and gravity arising in the linear
regime.  In terms of a rescaled time and a rescaled velocity measuring
departures from the Hubble flow, the adhesion model reduces to the
{\em Burgers equation} (originally, an equation for compressible
turbulence) \cite{adhesion,V-Frisch}:
\be
\frac{\partial \bm u}{\partial t} +
{\bm u}\cdot \nabla{\bm u} = \nu \nabla^2 {\bm u},
\label{Burgers}
\ee
with $\nabla \times {\bm u} = {\bm 0}$ by the parallelism assumption,
and the mass distribution is obtained from the field ${\bm u}({\bm
  r},t)$ by means of the continuity equation.
Here, $\nu \to 0^+$ is a phenomenological vanishingly small viscosity
modelling the coupling to the unresolved small--scale degrees of
freedom. The case $\nu=0$ corresponds actually to the Zel'dovich
approximation: the fluid elements move with constant velocity (in the
rescaled variables) along the initial gravitational acceleration, thus
effectively neglecting the effect of pressure, viscosity or other
small scale effects.  Obviously, singularities arise\footnote{The
situation is analogous to the formation of {\em caustics} in geometric
optics.} and the density field diverges. A small, but nonvanishing
value of $\nu$ regularizes these singularities into {\em shocks}: it
amounts to an {\em inelastic collision} prescription, such that fluid
particles adhere to each other at caustics, which become the walls
(``pancakes''), filaments and nodes that are typical of large scale
structure.

In the one-dimensional case, parallelism is exact.  With
scale-invariant initial conditions \cite{V-Frisch}, the mass
concentrates in shocks located in a {\em dense set} \footnote{The
adjective ``dense'' is understood with its mathematical meaning: a set
is dense in an interval, say, if in any sub-interval, however small,
there are points of the set.}, at which ${\bm u}$ has discontinuities.
This distribution is actually multifractal, namely, a peculiar type of
{\em bifractal}. Furthermore, this bifractal evolves in time: the
large shocks grow at the expense of smaller ones, illustrating the
bottom-up structure formation typical of cold dark matter.

In three dimensions, the walls, filaments and nodes are definite lower
dimensional objects, expected to arise in a generic situation.
Therefore, a {\em naive} picture of this structure consists of a
distribution of one, two and three-dimensional objects, that is, a
trivial example of multifractal distribution, with integer-dimension
objects only. However, as the initial velocity field is a non-smooth
Gaussian random field \cite{V-Frisch}, the structure produced
resembles a self-similar distribution of walls, filaments and nodes
that has been dubbed the {\em cosmic web}.  In this ``web'' the mass
concentrates, in addition to walls, filaments and nodes, in some
regions rather than in others (because those objects concentrate
there).  The cosmic web is a non-trivial multifractal.

\subsection{Kardar-Parisi-Zhang equation}
\label{KPZ}

The dissipative nature of the the Burgers equation (\ref{Burgers}) is
directly related to the coupling to small scales. It seems natural to
assume that the dissipation is complemented by a stochastic force (or
{\em noise}).  In terms of the velocity potential $\f$, i.e. ${\bm u}
= - \nabla\f$, the stochastic Burgers equation becomes the
Kardar-Parisi-Zhang equation\footnote{Under this name, this equation
is used for the description of surface growth.},
\begin{equation}
\fl
\frac{\partial \f}{\partial t} - \frac{1}{2}
(\nabla\f)^2 = \nu \nabla^2\f + \eta , \qquad
\langle \eta({\bm x},t)\,\eta({\bm x}',t')\rangle =
D\,\delta({\bm x}-{\bm x}') \,\delta(t-t') \,.
\label{KPZeq}
\end{equation}
The KPZ equation has critical regimes where $\f$ exhibits dynamical
scaling and the matter distribution is no longer determined solely by
the initial conditions, but is instead the outcome of the interplay
between the noise and the other terms of the equation.
As a general result, invariance under Galilean boosts required by the
convective nonlinearity
implies $\chi + z = 2$.

In the one-dimensional case, the large scale dynamics is dominated by
a weakly nonlinear (i.e., perturbative) fixed point at which $\chi =
1/2$ and $z = 3/2$. Furthermore, the Fokker-Planck equation for the
probability $\Pi[\f(x),t]$ associated to the KPZ equation
(\ref{KPZeq}) and its stationary solution are
\be
\fl
{\partial\Pi \over\partial t} =
-\int\! dx {\d \over\d \f(x)}[F(\f)\,\Pi] + {D\over 2}
\int\! dx {\d^2 \Pi\over\d \f(x)^2}
\label{1d-FP-KPZ}
\; \Rightarrow \;
\Pi_{\rm stat} =
\exp \left(-\frac{\nu}{2D}\int\!dx \,(\p_x\f)^2 \right),
\ee
where $F(\f) = \nu \,\p_x^2\f + \frac{1}{2} (\p_x\f)^2$.  $\Pi_{\rm
stat}$ is a Boltzmann velocity distribution at temperature $k T = 2
D/\nu$.  Although $\Pi_{\rm stat}$ is Gaussian, as in the absence of
nonlinearity, the value of $z$ is different from the linear-equation
value ($z_{\rm lin}=2$ associated to the diffusion equation). In
consequence, the dynamical scaling regime can be considered linear as
regards the stationary distribution, but not as regards temporal
scaling.  Some features involving temporal scaling can be calculated
perturbatively with the RG.  According to the stationary solution, the
shocks predicted by the adhesion model in $d=1$ disappear at large
times as the noise kicks in at a finite temperature $T$.

The critical dimension of the KPZ equation (\ref{KPZeq}) is $d=2$,
meaning that the nonlinearity is relevant if $d\leq 2$. For $d>2$, the
equation becomes perturbatively non-renormalizable, but there is
evidence of a scaling strong-coupling regime.  In particular, there is
evidence based on the non-perturbative RG \cite{Canet}, a tool that we
will introduce in Sect.\ \ref{ERG}.  In $d=3$, the KPZ equation driven
by the more general {\em coloured noise}, $\langle \eta({\bm
x},t)\,\eta({\bm x}',t')\rangle \propto |{\bm x}-{\bm x}'|^{2\rho-3}$,
has been considered in the cosmological context \cite{cosmoKPZ}.
There appear weakly nonlinear fixed points with exponents $\chi$, $z$
depending sensitively on the decay exponent $\rho$ of the noise
correlator.

A likely failure of the adhesion and KPZ approach is the absence of
vorticity imposed by the assumption of parallelism between velocity
and gravity.  Recently, Antonov \cite{Antonov} applied the RG to the
Burgers equation appended with a colored stochastic source of
vorticity. A perturbative fixed point arises where the scaling
behavior is related to the generation of vorticity and depends on the
decay exponent of the noise.

\section{Approaches based on coarse graining}

The KPZ models are rather phenomenological; namely, they miss a
first-principle derivation of or a physical argument for the choice of
noise correlator, which would make a quantitative prediction
available. 
Now we review briefly more systematic approaches, which are still
under development.

\subsection{The Small-Size Expansion}\label{}

The hydrodynamic equations are macroscopic equations, following from
microscopic Newtonian mechanics of particles through an averaging
process called {\em coarse graining}. This idea is implemented with
the help of a {\em window function} $W_L({\bm r})$, that is, a
function that quickly vanishes outside a neighborhood of the origin of
size $L$; typical examples are the sharp-cutoff (``top-hat'') window
and the Gaussian window. Thus, the coarse-grained mass density field
$\varrho_L({\bm r})$ is defined as the convolution
\be
\label{window}
\varrho_L({\bm r}) = \int d{\bm x} \;
W_L({\bm r} - {\bm x})\,\varrho({\bm x})
\ee
with the microscopic mass density field $\varrho({\bm x})$. One
defines similarly a velocity field ${\bm v}_L({\bm r})$, etc. The
exact set of equations for $\varrho_L$ and ${\bm v}_L$, corresponding
to the balance of mass and momentum, contain unknown terms which
describe the coupling of the coarse-grained dynamics to the degrees of
freedom in scales $<L$. Thus, a constitutive relation is required
expressing these terms as functions of the coarse-grained fields.

Recently, the ``Small-Size Expansion'' (SSE) has been proposed
\cite{SSE}, which relies on the physical hypothesis that the main
contribution by the small scales stems precisely from the scales close
to $L$. One finds a gradient expansion like those suggested in the
coarse-grain approach to incompressible turbulence.  In particular,
the phenomenological viscous term in the adhesion
model~(\ref{Burgers}) is replaced to lowest order in the expansion by
a term proportional to
\be
\frac{L^2}{\varrho_L} \left\{ (\nabla \varrho_L \cdot \nabla) {\bm g}_L
  - \nabla\cdot  \left[ \varrho_L \sum_k
    \left( \frac{\partial {\bm v}_L}{\partial x_k} \right)
    \left( \frac{\partial {\bm v}_L}{\partial x_k} \right)
  \right]
  \right\}
\ee
where ${\bm g}_L$ is the coarse-grained gravitational field. This term
may act like a drain of kinetic energy in collapsing regions, and the
adhesion model can be actually recovered under stronger dynamical
assumptions like parallelism of ${\bm v}_L$ and ${\bm g}_L$. However,
unlike the adhesion model, this term also behaves as a source of
vorticity.

\subsection{The exact renormalization group}
\label{ERG}

The coarse-grained variables change with the coarse-graining length
$L$ and so does the probability distribution, e.g., of the density
$\P_L[\varrho]$ (in Fourier space):
\be
\fl
{\partial\over \partial L}\P_L[\varrho] = 
{d \ln{\tilde W}_L^2\over
dL}\,{\delta \over \delta\varrho}(\varrho\,\P_L[\varrho]) +
{1\over 2}\,{d {\tilde W}_L^2\over
dL}\,P(k)\,{\delta^2 \over \delta\varrho^2}\P_L[\varrho]\,,
\label{ERGeq}
\ee
where ${\tilde W}_L(k)$ is the Fourier transform of the window
function, defined in Eq.~(\ref{window}), and $P(k)$ is the large-scale
power spectrum of the density fluctuations, i.e., the Fourier
transform of the correlation $\xi(r)$ introduced in Sec.\
\ref{sec:galaxies}.  The differential equation (\ref{ERGeq}),
describing evolution with the scale $L$, is the {\em exact RG}
equation for $\P_L$, which is a sort of Fokker-Planck equation for
$L$-evolution \cite{I-ERG}.

The exact RG equation has been amply used in high-energy and
 statistical physics \cite{WilKo,ERG}, but it has not been studied in
 the astrophysical literature (see \cite{I-ERG}).  However, it has
 been noticed that the one-point density probability distribution
 $p_L(\varrho)$ satisfies a diffusion equation which can be used to
 find the Press-Schechter mass function of collapsed objects
 \cite{Bond}.  This diffusion equation is connected with
 Eq.~(\ref{ERGeq}).

\subsubsection{Time evolution and RG}

Peebles \cite{Pee-RG} noted that the temporal variable, in some
cosmological solutions (scaling solutions), plays the role of a
scaling parameter and proposed a type of renormalization in which the
time evolution can be undone by a redefiniton of the space scale and
the number and mass of particles.  Therefore, Peebles assumes a
relationship between time evolution and evolution under the change of
scale, in the same fashion as dynamical scaling.  That relationship
admits a fuller formulation in terms of the exact RG:
Eq.~(\ref{ERGeq}) can be compared with a Fokker-Planck equation for
time evolution, like the one for the KPZ equation found in Sect.\
\ref{KPZ} \cite{I-ERG}.

\section{Conclusions}

From the observational standpoint, scaling in the large scale
structure is well justified, but the measures are not sufficiently
accurate yet to determine many details. The scale of homogeneity,
which has been the subject of much controversy, seems to be in the
range 10--20 Mpc/$h$.  There seems to be no point in trying to
determine a definite value of the fractal dimension. Rather, the
distribution fits a multifractal, so its scaling properties are given
by its multifractal spectrum, which can already be found with the help
of $N$-body simulations \cite{EPL}.  The determination of fine
morphological features requires further analysis of the galaxy
distribution as well as $N$-body simulations, with appropriate tools
such as statistical moments or more sophisticated tools (Minkowski
functionals, etc.). Among these morphological features are the voids.
Scaling of voids is beginning to be observed, but deeper studies of
voids will depend on improvement on their definition and, hence,
detection.

From the theoretical standpoint, scaling provides us with a handle in
an otherwise almost intractable problem of nonlinear dynamics. We have
seen that dynamical scaling is indeed sufficiently powerful to draw a
convincing picture of structure formation in one-dimension, with the
help of perturbation theory and the renormalization group.  However,
the realistic three-dimensional case may demand non-perturbative
tools, which are rather complex and are still being developed.

\end{document}